\documentclass[amsmath,amssymb,aps,floatfix,prl,reprint]{revtex4-1} 
\usepackage{graphicx}

\pdfoutput=1
\usepackage{amsmath}
\begin{document}
\title{Solvent Induced Proton Hopping at a Water-Oxide Interface}
\author{Gabriele Tocci}
\author{Angelos Michaelides}
\affiliation{Thomas Young Centre, London Centre for Nanotechnology and Department of Chemistry, University College London,
London WC1E 6BT, United Kingdom}
\email{angelos.michaelides@ucl.ac.uk}
\begin{center}
\begin{abstract}
Despite widespread interest, a detailed understanding of the dynamics of proton transfer at interfaces is lacking.
Here we use \textit{ab initio} molecular dynamics to unravel the connection between
interfacial water structure and proton transfer for the widely studied and experimentally well-characterized
water-ZnO(10$\bar{1}$0) interface. We find that upon going from a single layer of adsorbed water to a liquid multilayer
changes in the structure are accompanied by a dramatic increase in the proton transfer rate at the surface.
We show how hydrogen bonding and rather specific hydrogen bond fluctuations at the interface are responsible for
the change in the structure and proton transfer dynamics. The implications of this for the chemical reactivity
and for the modelling of complex wet oxide interfaces in general are also discussed.
\end{abstract}
\end{center}
\thispagestyle{empty}
\maketitle
Proton transfer in water is a process of central importance to a number of fields in science and technology.
Consider for example proton conduction across polymeric membranes
used in fuel cells~\cite{nature_fuel_cell_review} or through 
protein channels in cells~\cite{proton_transfer_water_chain_protein_2}. 
Proton transfer reactions are also key to many processes in catalysis such as the production
of hydrogen from methanol or biomass~\cite{nature_reforming_biomass,science_norsk_zno_cu_methanol},
or water formation~\cite{water_pt_angelos}. 
Whilst it is notoriously difficult to characterize
proton transfer under industrial or biological conditions, considerable insight and understanding has been gained
by examining well-defined model systems. One such model system is the example of the solvated proton in pure liquid water
~\cite{quantum_nature_proton_Tuckerman,marx_review,quantum_nature_proton_Marx_h3o+,tuckerman_prl09}.
Another model system is water adsorbed on atomically flat solid surfaces.
Indeed, whereas traditionally most work on well defined water/solid interfaces has focused on structure
characterization (e.g. Ref.~\cite{javi_review} and references therein),
increasingly the focus is turning to proton transfer and
related properties such as surface acidity and water dissociation~\cite{salmeron_nilsson_jacs_wat_copper,PRL_bonn,sulpizi_jctc,sprik_tio2_jctc,wat_gan_serra,wat_al2o3,wat_inp_gap_wood}.

Of the various water-solid interfaces that have been examined, water on ZnO$(10\bar{1}0)$
plays a central role in heterogeneous catalysis~\cite{science_norsk_zno_cu_methanol,MeyerPRLCuZnO,meyer_wat_zno04} 
and light harvesting~\cite{ZnO_dye}.
It is also a well-defined system that has
been the focus of a number of studies under
ultra high vacuum (UHV) conditions~\cite{meyer_wat_zno04,dulub_prl04}
which have hinted at potentially interesting dynamical behavior. 
Specifically, Meyer \textit{et al.} found that at monolayer (ML) coverage one out of every two water molecules is dissociated,
forming a so-called partially dissociated (PD) overlayer~\cite{meyer_wat_zno04}.
Subsequently they found that this PD overlayer could coexist with an overlayer of intact molecular (M) water~\cite{dulub_prl04}.
Moreover they suggested that the two states may rapidly interchange
such that an average configuration, intermediate between the two,
is at times observed in their scanning tunneling microscopy images.
These findings prompted a number of follow up studies which focussed on the structure of water 
on the surface or on the level of dissociation~\cite{woll_pccp,meyer_06pccp,raymand_2011,holtaus_jctc,todorova}.
This previous work indicates that
water on ZnO$(10\bar{1}0)$ might be a highly suitable system
for investigating proton hopping in interfacial water. However, the key issue of how proton hopping
occurs in this system and how it relates to the aqueous water environment is still not understood.
Indeed, this is true for most water/solid interfaces where major gaps in our
understanding of the mechanisms of proton motion at interfaces remain.

This work focuses on understanding proton transfer at the liquid water ZnO(10$\bar{1}$0) interface.
Although techniques for characterization of well-defined
aqueous interfaces have emerged (e.g. Refs.~\cite{salmeron_nilsson_jacs_wat_copper,PRL_bonn}),
probing the microscopic nature of proton transfer at interfaces remains a formidable challenge for experiment.
On the other hand, \textit{ab initio} molecular dynamics (AIMD), as we use here,
has reached such a state of maturity that it is now possible to simulate
bond making and bond breaking events at complex solid-liquid interfaces
(see e.g. Refs.~\cite{Galli_water_sic,wat_gan_serra,sprik_tio2_jctc,wat_nacl_pccp}).
Here, we find that upon going from a water ML -- characteristic of UHV --
to a liquid film (LF)  -- characteristic of ambient conditions -- changes
in the structure and in the proton transfer dynamics of interfacial water are observed.
Although moderate alterations in the structure of the contact layer are found,
the proton transfer rate increases more than tenfold.
Analysis reveals that H-bond fluctuations
induced by the liquid are responsible for the structural change
and for the substantial increase in proton transfer.
This effect is unlikely to be specific to water on ZnO, implying that
proton transfer may be significantly faster under aqueous conditions than at the low coverages
typical of UHV-style studies.
This fast proton transfer may also affect the chemical activity of a surface, being
particularly relevant to heterogeneous catalysis under wet conditions~\cite{hu_solvent,nature_zeol,neurock_zeol}.

The work reported here was carried out within the framework of density functional theory (DFT).
Full details of the computational set-up can be found in the Supporting Information~\cite{SI}. 
However, in brief, the key features of the simulations are that we used the PBE~\cite{pbe}
exchange-correlation functional  and the CP2K code~\cite{cp2kvond}.
The surface model is made of $6\times 3$ primitive surface unit cells and a 3 bilayer slab.
There is one water molecule per primitive cell at ML coverage,
whereas the LF is comprised of 144 molecules, resulting in a $\approx 2$ nm thick overlayer.
The AIMD simulations are performed in the canonical ensemble  close to room temperature.
We performed extensive tests on the set-up to explore the sensitivity of our results to issues such
as basis set and exchange-correlation functional, including functionals that account for exact exchange and van der Waals forces~\cite{SI}.
Overall we find that compared to other interfacial water systems
this one is rather benign and none of our main conclusions are affected
by the specific details of the DFT set-up.
In particular, although the importance of van der Waals dispersion forces between water molecules
and water on surfaces is being increasingly recognised
(see e.g. Refs.~\cite{jiri_rev,bis_prl_ice,Galli_water_vdw,javi_vdw_prl}) they do not have a significant impact
on the dynamics of this system~\cite{SI}.
\begin{figure}[!b]
{\includegraphics[width=0.5\textwidth]{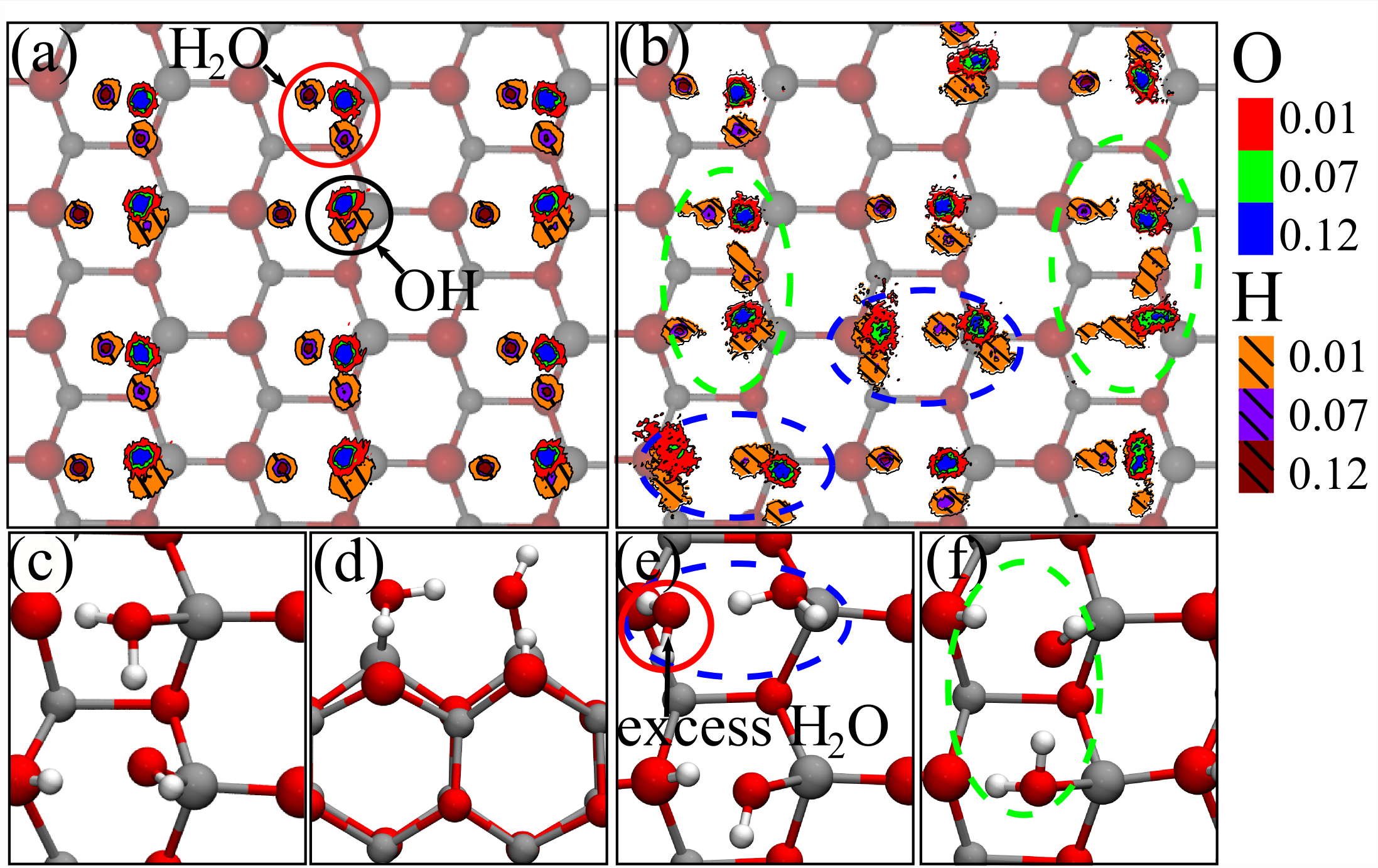}
\caption{Spatial probability distribution function of the O and H atoms
projected on ZnO$(10\bar{1}0)$ for (a) the water monolayer and (b) the contact layer of the liquid film.
Gray, red and white spheres are Zn O, and H atoms, respectively.
The topmost Zn and O surface atoms are shown using larger spheres.
In (a) a H$_2$O and a OH that are connected 
via a H-bond are circled in red and black, respectively.
(c) top and (d) side view of the partially dissociated water dimer,
which is the basic building block of the $(2\times 1)$ overlayer structure.
Snapshots of the liquid film showing water in a new type of structure enclosed in a blue oval (e) and
partially dissociated dimer structure enclosed in a green oval (f).}
\label{fig:spat_dist}}
\end{figure}

Let us first consider the adsorption of water on ZnO$(10\bar{1}0)$ at ML coverage.
Fig.~\ref{fig:spat_dist}(a) shows the spatial probability distribution
function of the O and H atoms adsorbed on the surface at ML
coverage. This illustrates the average structure of the
overlayer projected onto the surface.
Only the PD structure is observed, and it has a similar structure
(bond lengths differ by $<0.05$ {\AA}) to the zero temperature geometry optimized structure.
Figs.~\ref{fig:spat_dist}(c) and (d) show snapshots of the PD state
in top and side views, respectively.
The OHs and the H$_2$Os sit in the
trenches and are covalently bound to the surface-Zn atoms.
A H-bond is formed between the surface-Os and the H$_2$Os and also
between the surface OHs and the dissociated water.
In addition, the H$_2$Os donate a H-bond to the neighbouring OHs and
lie essentially flat on the surface, whereas the OHs are tilted up and point away from the surface.

\begin{figure}[!t]
{\includegraphics[width=0.5\textwidth]{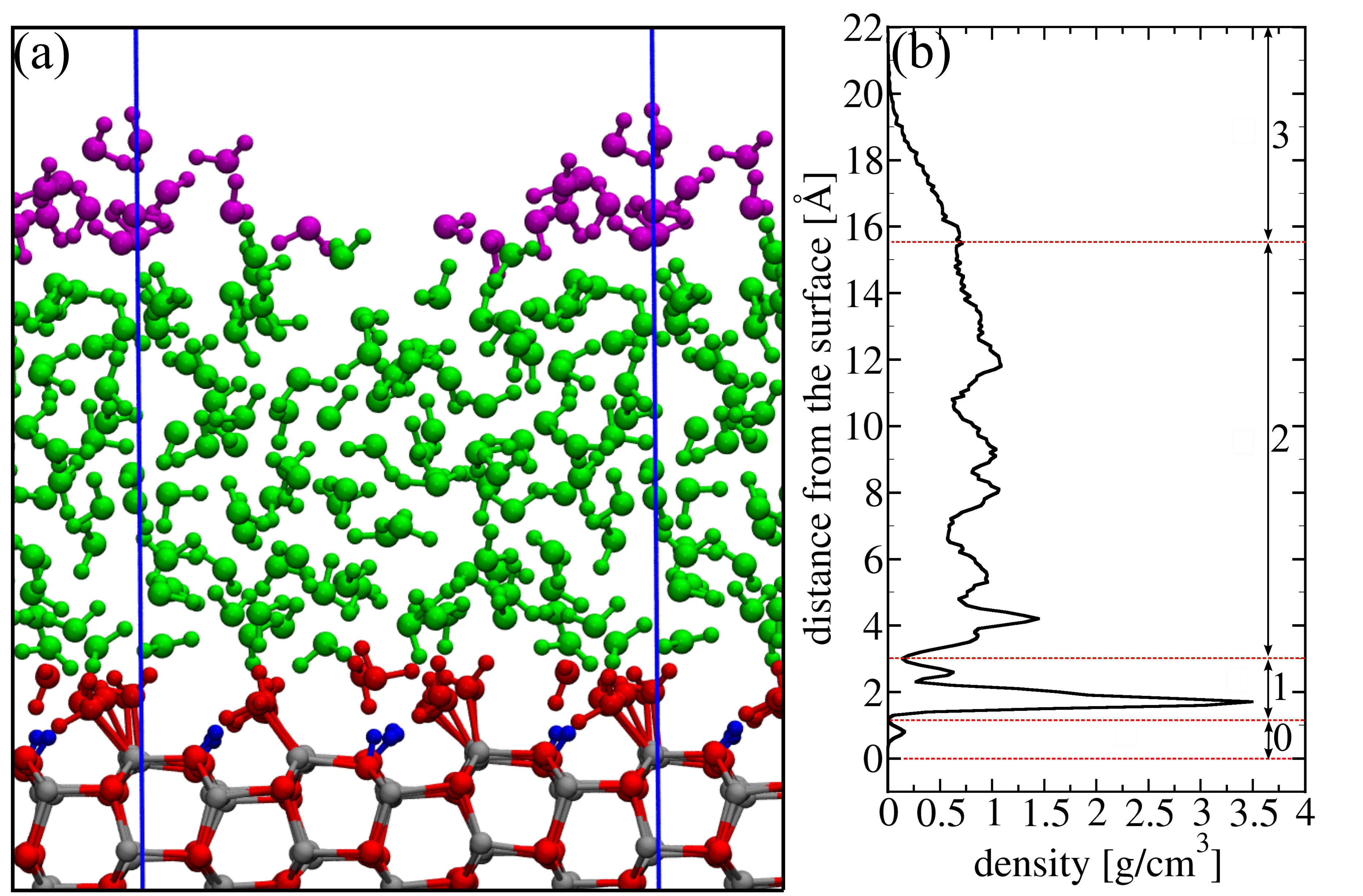}
\caption{(a) Snapshot of a liquid water film on ZnO$(10\bar{1}0)$(a).
(b) Planar averaged density profile as a function of the distance
from the surface, where different regions are
identified and labelled from 0 to 3. In (b)
the zero in the distance is the average height of the top surface ZnO layer
and the density reported is the planar averaged density of adsorbed species.
In (a) the top four surface layers 
are shown and the water overlayer is colored according to the regions shown
in the density profile (b).  Regions  (going from 0 to 3) correspond
to chemisorbed H atoms, H$_2$Os/OHs adsorbed on the surface,
mainly bulk-like liquid water, and water in the liquid vapor interface.
}
\label{fig:str}}
\end{figure}

A snapshot of the liquid water film is illustrated in Fig.~\ref{fig:str}(a) and in
Fig.~\ref{fig:str}(b) the planar averaged density profile as a function of distance from the
surface is reported. The density profile shows a pronounced layering,
as previously reported for water on various
substrates~\cite{wat_inp_gap_wood,wat_nacl,Galli_water_graphene,fenter04_prog_surf_sci,PhysRevLett_wat_mica}.
For convenience we discuss  the density profile in terms of
the regions observed, and label them from 0 to 3.
Region 0  shows up as a small peak close to the
surface and this corresponds to the chemisorbed Hs.
These are the Hs that bond to the surface as a result of
dissociation of some of the H$_2$Os.
The large peak of $\approx 3.2$ g/cm$^3$ in region 1 at about 2.0 {\AA} corresponds to a mixture of OHs
and H$_2$Os in immediate contact with the surface. The second peak in region 1 of about 0.7 g/cm$^3$
also arises from a mixture of OHs and H$_2$Os,
that however sit on top of the surface-O atom. 
Between regions 1 and 2 there is a depletion of H$_2$Os, then in region 2 the oscillations are damped
until in region 3 the density decay, characteristic of the liquid-vacuum interface, is observed~\cite{wat_kuo_sci}.

\begin{figure}[!b]
{\includegraphics[width=0.5\textwidth]{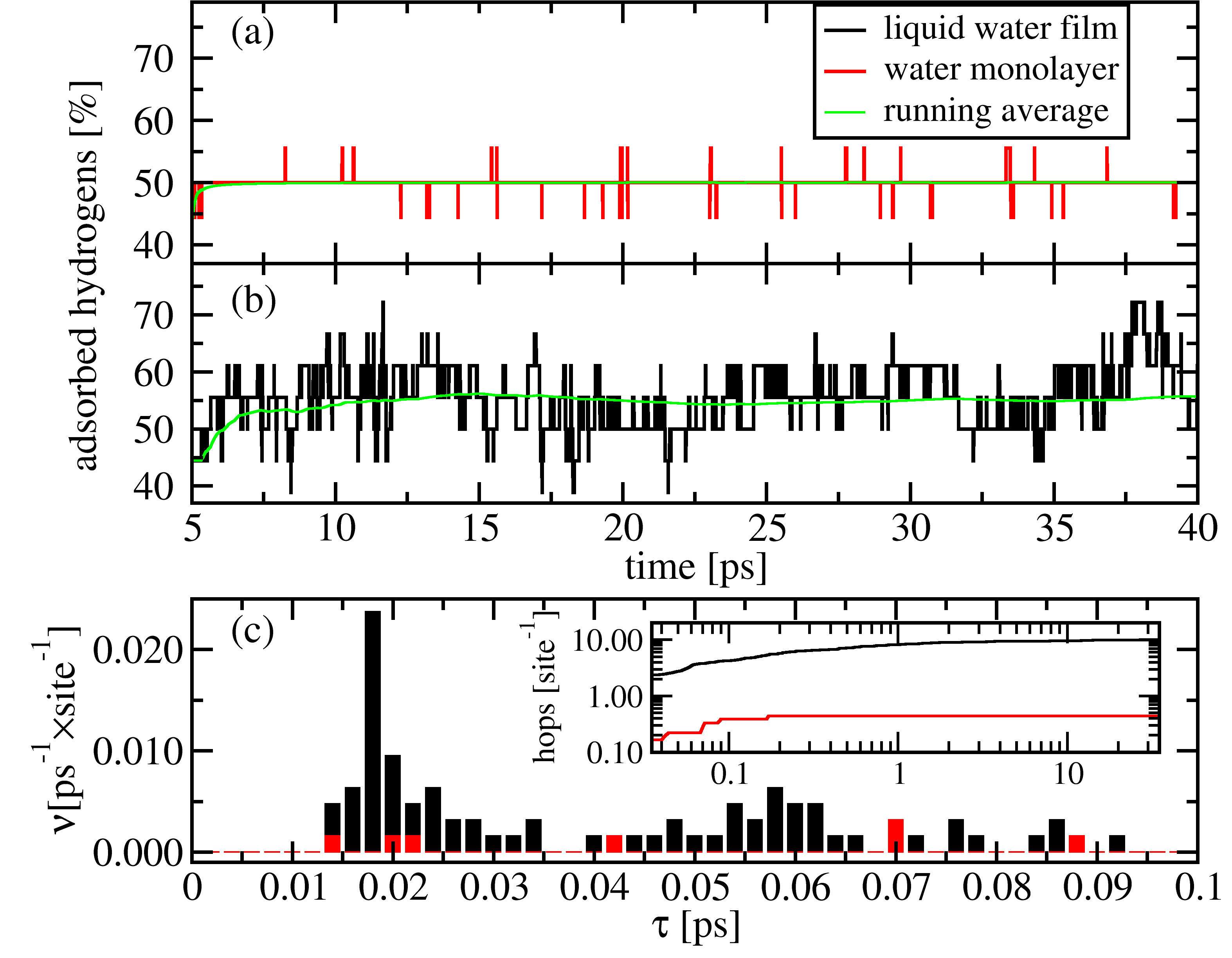}
\caption{Time evolution of the percentage of H atoms adsorbed on the surface
for (a) water monolayer and (b) liquid water overlayer.
(c) Proton hopping frequency $\nu(\tau)$ as a function of the
residence time $\tau$ for the liquid film (black)
and the monolayer (red).
The inset is a $\log$-$\log$ plot of the total number of hops as a function of $\tau$,
obtained as $\int_0^\tau \nu(\tau^{\prime}) d \tau^{\prime} $.
The full 35 ps of analysis are shown in the inset.
\label{fig:ads_Hs}}}
\end{figure}

The structure of the contact layer in the LF differs from the ML in a
number of ways (\textit{c.f}. Figs.~\ref{fig:spat_dist}(a) and (b)).
First, although there are remnants of the
($2\times 1$) structure (see green ovals in Figs.~\ref{fig:spat_dist}(b) and (f)), the symmetry
present at ML coverage is now broken.
Second, the proton distribution is more delocalized
in the contact layer of the LF than in the ML.
Third, and most notably, the coverage in the LF has
increased to $1.16\pm 0.03$, with excess waters sitting
in a new configuration on top of a surface-O
(circled in red in Fig.~\ref{fig:spat_dist}(e)). 
At this new site adsorption can happen
either molecularly or dissociatively and in either case the adsorbate accepts a H-bond
from a H$_2$O sitting on the top-Zn site. 
Analysis of the overlayer reveals that 
H-bonding with the liquid above stabilizes the excess H$_2$O at the
top-O site which gives rise to the higher coverage~\cite{note_diff_ads_energy}.
Despite the structural change  between the ML and the contact layer of the LF,
we did not observe any exchange of water. Further,
the level of dissociation is not altered in the two cases.
This can be seen in Figs.~\ref{fig:ads_Hs}(a) and (b) where
the trajectory of the percentage of adsorbed H atoms is reported for the two systems.
At an average of $50\%$ dissociation in the case of the ML and  $55 \pm 5 \%$
for the contact layer of the LF the difference is not significant~\cite{disagreement_reaxff}.

Whilst the changes in the structure between the ML and LF
are interesting and important, remarkable differences in the proton transfer dynamics are
observed. This is partly shown by the fluctuations in the percentage of adsorbed H atoms,
which represent proton transfer events to and from the surface (Fig. 3).
Clearly by comparing Figs.~\ref{fig:ads_Hs}(a) and (b) it can be seen that
the fluctuations are much more pronounced in the LF than in the ML.
However, proton transfer to and from the surface is only part of the story
as proton hopping between the H$_2$Os and OHs is also observed in the contact layer of the LF.
Indeed this is already clear by looking at the proton distribution within the green ovals
in Fig.~\ref{fig:spat_dist}(b). In the analysis reported in Fig.~\ref{fig:ads_Hs}(c) all events are included  and the hopping of
each proton is monitored. Specifically, we plot the hopping frequency ($\nu=$ number of hops$/$(time$\times$sites)) against $\tau$.
$\tau$ is defined as the time a proton takes to return to the O it was initially bonded to,
and therefore measures the lifetime of a proton transfer event, with larger values of $\tau$ corresponding to longer lived events.
Fig.~\ref{fig:ads_Hs}(c) thus reveals that proton transfer is more frequent in the LF than in the ML.
Specifically, in the LF there are more events at all values of $\tau$, with a maximum in the frequency distribution
of about  0.02$/$(ps$\times$site) at $\tau \approx 20$ fs.
In contrast in the ML the $\nu$ distribution never reaches values larger than 0.005$/$(ps$\times$site).
The $\approx 20$ fs lifetime of the hopping events observed here is similar to the timescale of
interconversion between Zundel-like and Eigen-like complexes in liquid water ($< 100$ fs) obtained
from femtosecond spectroscopy. It is also in the same ballpark as other theoretical estimates of proton transfer lifetimes obtained from
work on proton transport in  liquid water or on other
water/solid interfaces~\cite{tuckerman_prl09,fs_spectroscopy_liq_wat,jacs_akimov_2013}.
The total number of hops (inset in Fig.~\ref{fig:ads_Hs}(c)) is $\approx 0.4/$site
but about $10/$site in the LF.
While only proton hopping between the overlayer and the surface is observed in the ML,
in the LF $\approx 1/4$ of the hops are within the contact layer with the remaining $3/4$ of all hops being
to and from the surface. Proton hopping events are also longer lived in the LF than in the ML.
This is demonstrated by the long tail in the frequency distribution of the
LF and more clearly by the inset in Fig.~\ref{fig:ads_Hs}(c),
which shows that the longest hopping events are only about 0.2 ps in the ML but as long as $\approx 4$ ps in the LF.
Events with a lifetime of the order of the picosecond are characteristic of Grotthus-like diffusion~\cite{marx_review}
in liquid water or in other water/solid interfaces~\cite{jacs_akimov_2013,wat_inp_gap_wood,wat_inp_gap_wood_jpcc,wat_gan_serra},
which are however not observed here, although may occur at longer timescales than we can simulate~\cite{raymand_2011}.
\begin{figure}[!t]
{\includegraphics[width=0.5 \textwidth]{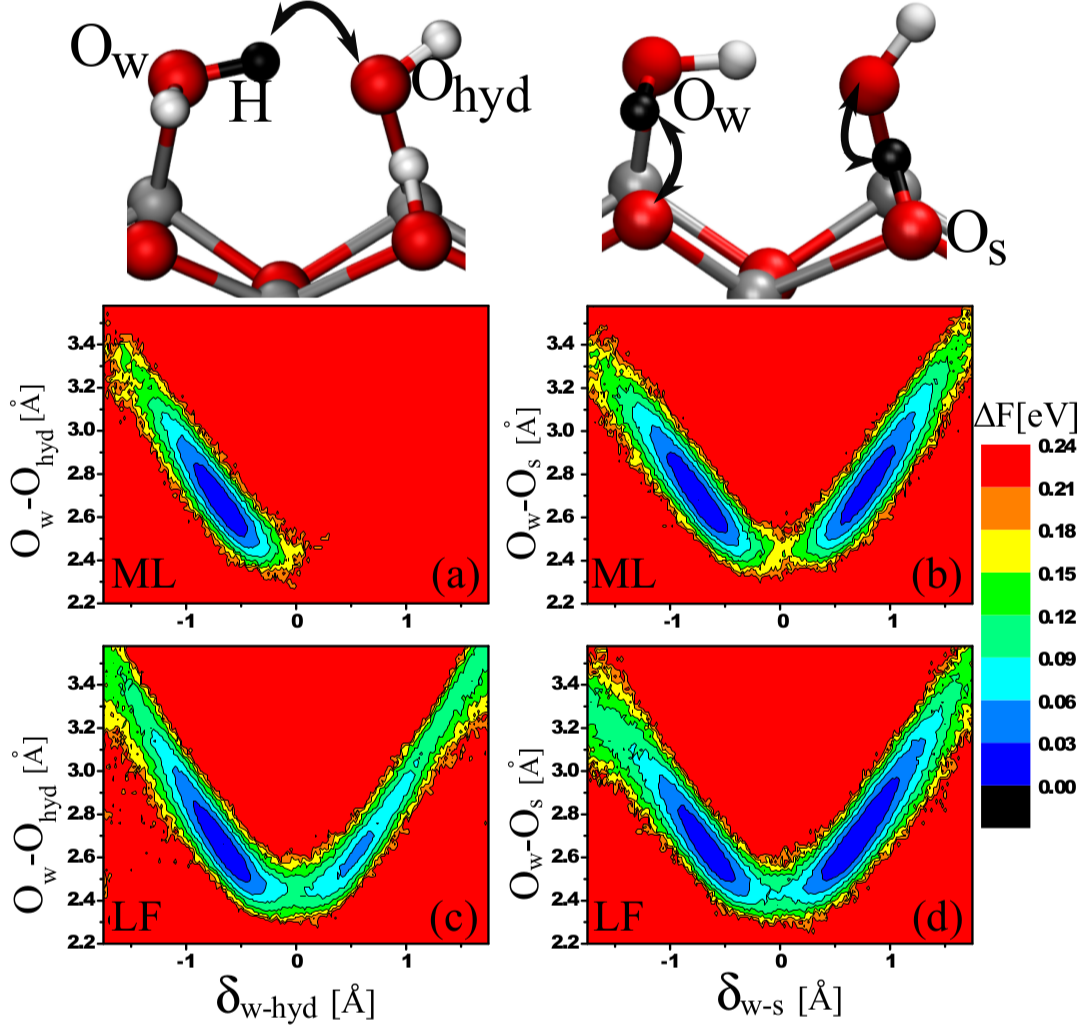}
\caption{Free energy, $\Delta F$ contour plots for protons hopping between two Os
as a function of the O-O distance and of the location of the protons between the two Os, $\delta$.
(a) and (b) show the free energies for the protons hopping in the water monolayer (ML) simulation, while (c) and (d) illustrate free energies
in the liquid film (LF) simulation. As illustrated by the structures at the top of the figure,
(a) and (c) refer to hopping between the Os in the contact layer and
(b) and (d) refer to hopping of protons to and from the surface.
The contour lines and colours are shown on the same scales.
\label{fig:pdf_delta}}}
\end{figure}

To gain further insights in the two systems
we plot in Fig.~\ref{fig:pdf_delta} free energy ($\Delta F$) surfaces $\Delta F$ for the
various distinct proton transfer events considered here. The free energy surfaces
have been obtained in a standard manner from $\Delta F= -k_B T \log P($O-O,$\delta)$.
The probability distribution $P($O-O,$\delta)$ is  a function
of O-O distances and of $\delta$, the position of the H with respect to the two Os.
With reference to Fig.~\ref{fig:pdf_delta} $\delta_{1\textrm{-}2}$
is defined as the difference in the distances between H and two oxygens, O$_1$ and O$_2$, i.e.
$\delta_{1\textrm{-}2}=$ O$_{1}$-H $-$ O$_{2}$-H.
There are some clear differences between the free energy maps of the ML and the LF. First, the single
minimum in Fig.~\ref{fig:pdf_delta}(a) shows that in the ML
protons do not hop between adsorbed H$_2$O and OHs. In contrast in the LF two clear minima
are identified revealing that hopping between adsorbed
H$_2$Os and OHs occurs readily. The approximate free energy barrier
is $\approx 100$ meV. Secondly, proton hopping to and from the surface
happens both in the ML (Fig.~\ref{fig:pdf_delta}(b) and in the LF (Fig.\ref{fig:pdf_delta}(d)),
but the free energy barrier is noticeably lower in the LF ($\approx 70$ meV) than it is for the ML ($\approx 160$ meV).

\begin{figure}[!h]
{\includegraphics[width=0.5\textwidth]{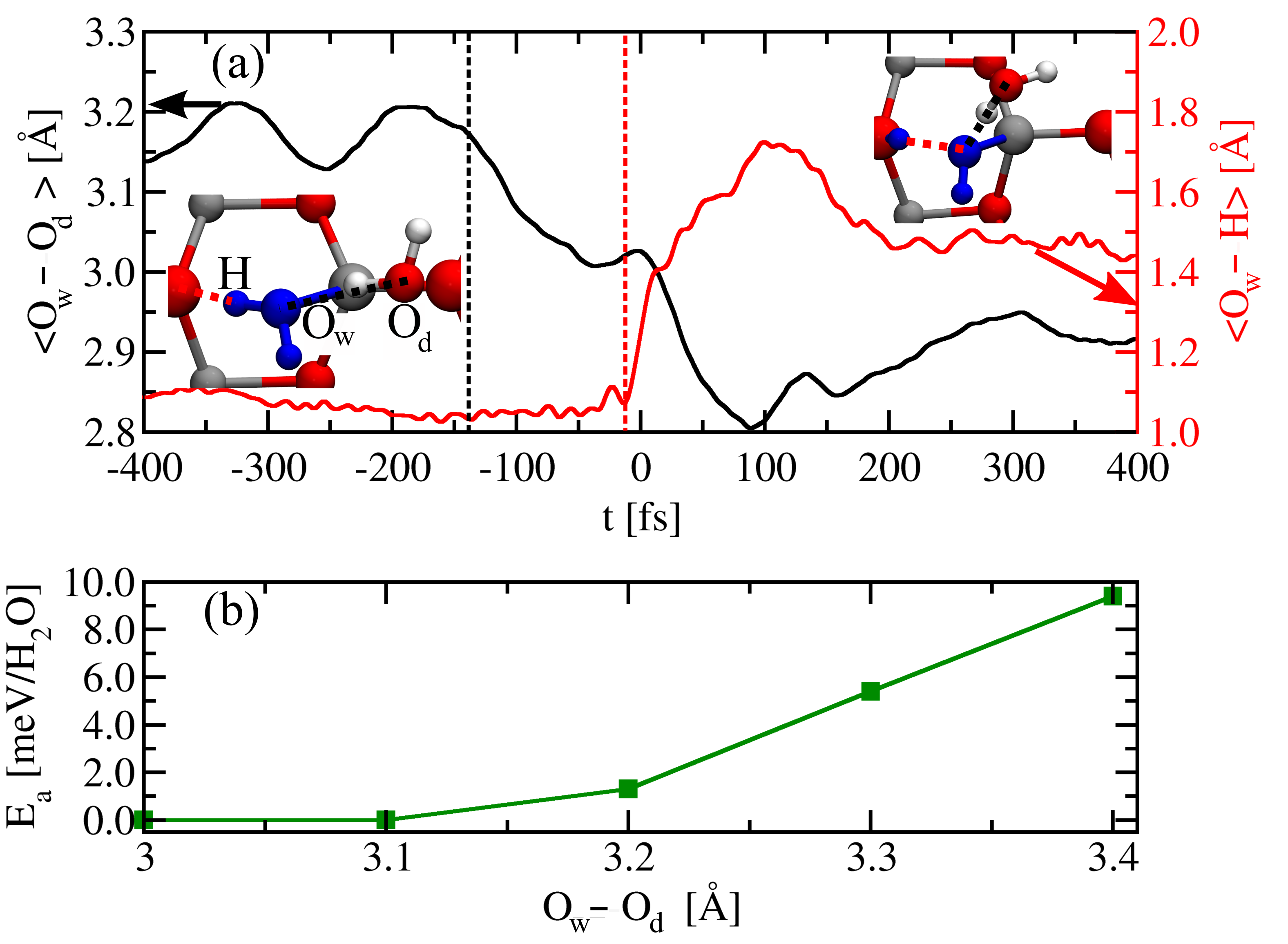}
\caption{Analysis of the role of H-bond fluctuations on proton transfer. (a) Average O-O distance and
average O-H distance as a function of time for all proton transfer events to the surface.
The O-O distance plotted (black line) is the distance between the O
of the molecule involved in the proton transfer event (O$_{\mathrm{w}}$)
and the O of the nearest molecule from which it accepts a H-bond (O$_{\mathrm{d}}$).
The O-H distance (red line) is the distance between O$_{\mathrm{d}}$  and the H that is involved in proton transfer.
The black and red vertical lines indicate the approximate moment where there is a significant
change in the $\langle$O$_{\mathrm{w}}$-O$_{\mathrm{d}}\rangle$  distances, respectively.
The insets show snapshots of specific molecules before and after dissociation.
(b) Activation energy (E$_a$) for water dissociation at ML coverage as a
function of the O$_{\mathrm{w}}$-O$_{\mathrm{d}}$ distance (calculated using VASP~\cite{vasp1,vasp2}, see~\cite{SI}).
\label{fig:barrier_ML_constr}}}
\end{figure}

In order to understand why the hopping frequency increases so much upon going from ML
to multilayer we have examined the time dependence of the H-bonding network at the interface.
This reveals an intimate connection between the local H-bonding environment of a molecule and its proclivity towards proton transfer.
From the AIMD trajectory we see this connection between H-bonding environment and the hopping of
individual protons and we demonstrate in Fig.~\ref{fig:barrier_ML_constr}(a)
that this holds on average for the entirety of all water-to-surface proton hopping events.
Specifically, Fig.~\ref{fig:barrier_ML_constr}(a) shows the mean length
of the O-H bonds that break in a proton transfer event ($\langle$O$_{\mathrm{w}}$-H$\rangle$) as a function of time.
We find that this is correlated with $\langle$O$_{\mathrm{w}}$-O$_{\mathrm{d}}\rangle$, the mean distance between O$_{\mathrm{w}}$ and
O$_{\mathrm{d}}$, where O$_{\mathrm{d}}$ is the O of the nearest molecule donating a H-bond to O$_{\mathrm{w}}$.
At time t$\,<0$ water is intact at a distance $\langle$O$_{\mathrm{w}}$-H$\rangle \approx 1.0$ {\AA}.
Just before $t=0$, the point at which the $\langle$O$_{\mathrm{w}}$-H$\rangle$ bond breaks, there is a sharp increase in the
$\langle$O$_{\mathrm{w}}$-H$\rangle$ distance and then it levels off at
$\approx$ 1.4 {\AA}, about 200 fs after  dissociation.
Accompanying these changes in the $\langle$O$_{\mathrm{w}}$-H$\rangle$ distance are changes in
$\langle$O$_{\mathrm{w}}$-O$_{\mathrm{d}}\rangle$ distances.
Crucially about $150$ fs before proton transfer there is a net decrease in the intermolecular separation
that shortens $\langle$O$_{\mathrm{w}}$-O$_{\mathrm{d}}\rangle$ from about 3.1 to 2.9 {\AA}.
It can be seen clearly that this change in intermolecular separation occurs before the
$\langle$O$_{\mathrm{w}}$-H$\rangle$ bonds start to break revealing that rearrangement in H-bonding is required prior to proton transfer.
Similar behaviour has recently been reported for the
liquid water/InP(001) interface~\cite{wat_inp_gap_wood}.
Further, O-H bond lengthening due to the presence of additional
water was reported for water on Al$_2$O$_3$~\cite{wat_al2o3}.
Here, we illustrate that an increase in the O-H bond length occurs before the O-O distance decreases.
Not only are the two distances correlated but it is the decrease in the O-O distance
which produces an increase in the O-H distance.

Through a careful series of additional calculations in which an individual proton transfer event was examined
we established that the proton transfer barrier depends critically on the intermolecular distance.
As shown in Fig.~\ref{fig:barrier_ML_constr}(b) for 
relatively large distances of 3.4 {\AA} there is a small $\approx 10$ meV barrier. As the O$_{\mathrm{w}}$-O$_{\mathrm{d}}$ 
distance decreases, so too does the barrier until at $ \approx 3.1$ {\AA}
where there is no barrier and the intact water state is not stable.
Given that fluctuations in the H-bond distances are more pronounced in the LF than in the ML
and lead at times to relatively short O$_{\mathrm{w}}$-O$_{\mathrm{d}}$ separations,
it is this that causes the more frequent proton transfer.
An estimate of the H-bond distance fluctuations is obtained by computing the root mean square displacement
the O-O distances in the contact layer, which gives 0.43 {\AA} in the LF compared to the much smaller value of 0.15 {\AA} in the ML.
This increase is also the reason why hopping does not occur between neighbouring H$_2$Os and OHs on the ML while it
does in the LF. H-bond distance fluctuations are also responsible for a proportion of events having
a long lifetime (with $\tau \approx 1$ ps, see inset of Fig.~\ref{fig:ads_Hs}(c)),
although actual hydrogen bond forming and breaking and not just fluctuations in the distance may participate in this case.
While we never observe H-bond forming or breaking in the ML, the H-bond lifetime is of the order of the picosecond in the LF
and this correlates well with the long lived proton transfer events.

We have shown that there are clear differences in the properties
of water in contact with ZnO$(10\bar{1}0)$ upon going from UHV-like
to more ambient-like conditions.
Changes in the adsorption structure upon increasing the coverage above 1 ML 
have previously been predicted for  a number of
substrates including ZnO$(10\bar{1}0)$~\cite{wat_tio2,wat_tio2_reply,wat_al2o3,raymand_2011,salmeron_nilsson_jacs_wat_copper}.
The specific observation here is that the liquid water film leads to
a $\approx 16\%$ increase in the water coverage and
a breaking of the $(2\times 1)$ periodicity observed at ML.
This arises because of H-bonding between the molecules in the contact layer and the molecules above it.
It should be possible to verify this increased capacity for water adsorption
using a technique such as \textit{in situ} surface X-ray diffraction.

We have demonstrated that there is
a substantial increase in the proton transfer rate in the contact layer of the LF
and that this is caused by H-bond fluctuations that lower the proton transfer barrier.
A H-bond induced lowering of the dissociation barrier upon increasing the water coverage has been discussed
before~\cite{water_diss_ru_mich,meyer_wat_zno04,wat_mgo_odel,salmeron_nilsson_jacs_wat_copper,xiaoliang_wat_mgo,wat_al2o3,wat_inp_gap_wood_jpcc}.
 Here, however, we have demonstrated that the barriers to dissociation and recombination are lowered
in general because of the presence of the liquid.
As in the case of liquid water and water on other substrates we show (Fig.~\ref{fig:pdf_delta})
that there is a strong dependence between the proton transfer barrier and the distance
between the Os on either side of the hopping proton~\cite{quantum_nature_proton_Tuckerman,marx_review,xin_wat_metal,wat_gan_serra}.
However, we have also identified a connection between the molecule involved in
the proton transfer and the molecules in its first solvation shell (Fig.~\ref{fig:barrier_ML_constr}).
This observation is somewhat similar to the structural diffusion of the excess proton in
liquid water~\cite{marx_review,quantum_nature_proton_Marx_h3o+}.
The key difference between the two is that concerted H-bond breaking and making is required for proton diffusion
in liquid water~\cite{tuckerman_prl09}, while only fluctuations in the H-bond distance are needed for proton
transfer (but not diffusion) to occur. Because fluctuations of the solvent provide the mechanism for the increased proton
transfer rate, a similar effect is expected also on other substrates, e.g. on reactive metal surfaces upon which water dissociates~\cite{water_diss_ru_mich,salmeron_nilsson_jacs_wat_copper}.

Finally, since the barrier to proton transfer is sensitive to changes
in specific H-bond distances it is likely that
implicit solvent models will be inadequate for this class of
system as they do not account for H-bond fluctuations.
A solvent induced increase in the proton transfer rate may also affect the
chemical activity of the substrate and therefore have important consequences
for heterogeneous catalysis under wet
conditions~\cite{hu_solvent,nature_zeol,neurock_zeol,science_FeO_Merte}.
Given that the O-O distance correlates with the barrier height and that H-bond distances of adsorbed H$_2$Os/OHs
are related to the lattice constant of the substrate, it might be possible to tailor
the proton hopping rate through e.g. strain or doping of the substrate. 

In conclusion, we have reported on a detailed AIMD
study of water on ZnO. In so doing we have tried to bridge the gap between studies of proton transfer
in liquid water and low coverage UHV-style work.
This has revealed a substantial increase in the rate of proton transfer upon increasing
the coverage from a monolayer to a liquid multilayer.
We have tracked down the enhanced proton transfer rate to specific solvent induced fluctuations
in the H-bond network, which yield configurations with
relatively short intermolecular distances wherein the barrier to proton transfer is
lowered. These findings are potentially relevant to the modelling of wet interfaces in general
and to heterogeneous catalysis.
\section{Supplementary Information}
Further tests and computational details. This material
is available free of charge via the Internet at DOI:\href{http://pubs.acs.org/doi/abs/10.1021/jz402646c#10.1021/jz402646c}{10.1021/jz402646c}
\section{Acknowledgments}
We are grateful for support from the FP7 Marie Curie Actions of 
the European Commission, via the Initial Training Network SMALL (MCITN-238804).
A. M. is supported by the European Research Council and the Royal Society
through a Wolfson Research Merit Award.
We are also grateful for computational resources to the London Centre for
Nanotechnology and to the UK's HPC Materials Chemistry
Consortium, which is funded by EPSRC (EP/F067496), for access to HECToR.
\bibliography{water_on_zno_bib}

\begin{thebibliography}{10}
\bibitem{note_diff_ads_energy}%
  \BibitemOpen
  \bibinfo {note} {From a separate set of geometry optimizations we established
  that the contact layer of the LF (without the liquid overlayer above it) is
  90 $\pm \, 30$ meV/H$_2$O less stable than the structure of the ML. This
  value was obtained by comparing the energy of the optimized ML with the
  average energy of 5 characteristic snapshots of the contact layer from the LF
  simulation (each relaxed in the absence of the liquid overlayer).}%
  \bibAnnoteFile{Stop}{note_diff_ads_energy}%
\end{thebibliography}
\end{document}